\documentstyle[12pt,aaspp4,psfig]{article}

\lefthead{left}
\righthead{right}

\newcommand{\eps}{\varepsilon}
\newcommand{\zmax}{z_{\rm max}} 
\newcommand{\gtilde}
 {~ \raisebox{-1ex}{$\stackrel{\textstyle >}{\sim}$} ~}
\newcommand{\ltilde}
 {~ \raisebox{-1ex}{$\stackrel{\textstyle <}{\sim}$} ~}

\begin{document}

\footnotesize

\pretolerance=100
\tolerance=200
\rightskip=0pt

\title{Probing the Cosmic Star Formation History by Brightness Distribution
of Gamma-Ray Bursts}

\author{Tomonori Totani}
\affil{Department of Physics, School of Science,
The University of Tokyo, Tokyo 113-0033,
Japan \\ E-mail: totani@utaphp2.phys.s.u-tokyo.ac.jp}

\begin{center}
\it Accepted by Astrophysical Journal
\end{center}

\begin{abstract}
Brightness distribution of Gamma-Ray Bursts (GRBs) is studied in detail
under the assumption that GRB rate is related to cosmic star formation rate.
The two populations of the long- and short-duration
bursts in the 4B BATSE catalog are analyzed 
separately. Taking account of current uncertainties
in the observational estimate of star formation rate (SFR), we have tried
various models of the cosmic star formation history  and
we find that the SFR evolution in $z$ = 0--1 is strongly constrained
by the GRB distribution if the standard candle approximation is valid.
The strong SFR evolution by a factor of $\sim$ 15
from $z$ = 0 to 1 inferred from UV 
observations is too steep to be consistent with the GRB distribution
for any distance scale of GRBs.
Some possibilities to reconcile this discrepancy are discussed, 
including the intrinsic luminosity dispersion of GRBs and/or modification of 
star formation history estimated by UV observations. 
We argue that SFR increase factor from $z$ = 0 to 1 may be as low as about 4
if we choose different sets of cosmological parameters and/or
take account of the evolution of metallicity and dust extinction in
the UV data, and this would significantly remedy the discrepancy.
\end{abstract}


\section{Introduction}
The brightness distribution of gamma-ray bursts (GRBs) 
observed by the Burst and Transient Source Experiment (BATSE)
has been known
to be significantly deficient in faint bursts compared with that expected in
the Euclidean space (Meegan et al. 1992), and this
has been considered as one of the evidences of the
cosmological origin of GRBs (Mao \& Paczy\'{n}ski 1992; Piran 1992; 
Dermer 1992). On the other hand, since most of GRB models
are associated to death of massive stars whose lifetime is much
shorter than the cosmological time scale, the brightness distribution
of GRBs reflects not only the cosmological effects but also 
the cosmic star formation history (Totani 1997, hereafter T97; 
Sahu et al. 1997; Wijers et al. 1998). The cosmological origin of GRBs
is now confirmed by the discovery of metal absorption lines
in the optical counterpart of GRB 970508 (Metzger et al. 1997),
and hence
more detailed analyses of the GRB brightness distribution 
are required to investigate the
origin of GRBs and the cosmic star formation history.

The most important consequence of the possible relation of GRB rate and
star formation rate (SFR) is that
the distance scale and absolute luminosity of GRBs become
larger than those for no-evolution sources, because we need stronger
cosmological effect in order to cancel out the rapid increase of
the observed star formation rate 
from $z=0$ to 1 by a factor of more than 10 (Lilly et al. 1996). 
However, different authors
give quantitatively different results: T97 pointed out that 
the scenario of binary neutron-star mergers (Blinnikov et al. 1984,
referred to ``the NS-NS model'', hereafter) results
in a better fit than the case that GRB rate is simply proportional
to SFR (referred to ``the proportional model'', hereafter),
because the time delay during the spiral-in phase of binary
neutron stars make the SFR evolution significantly
flatter in $z$ = 0--1. On the other hand,
Wijers et al. (1998) concluded that the proportional model
is in good agreement with the observed GRB brightness distribution
when the redshift of the most distant GRBs is about 6.
Petrosian \& Lloyd (1997), however, claimed that neither 
the NS-NS nor the proportional model is consistent with
the observationally determined SFR evolution. This paper investigates
the origin of this discrepancy paying special attention to uncertainties
in the SFR observations. We found that the GRB rate evolution 
in $z$ = 0--1 is crucially important in the GRB distribution analysis,
and constraints on the cosmic star formation history
in this redshift range are obtained, under the assumption of 
the standard candle approximation.

It is well known that the duration of GRBs shows a bimodal distribution
suggesting the existence of two populations of long ($\gtilde$2 sec) and
short ($\ltilde$ 2 sec) GRBs (Kouveliotou et al. 1993).  
All previous papers including T97,
which investigated the GRB brightness distribution in the context of
the cosmic star formation history, however treated the GRBs as a
single population. Because the energy spectrum of short GRBs is
significantly harder than that of long GRBs (Kouveliotou et al. 1993)
and the assumed spectrum affects
the GRB brightness distribution analysis,
we should analyze the GRB distribution separating the two populations.
We define the long GRBs as those with $T_{90} >$ 2 sec,
while the short GRBs with $T_{90} <$ 2 sec, where $T_{90}$ is the interval 
over which 5 \% to 95 \% of the burst counts accumulate (Kouveliotou et al. 
1993), and analyze them separately in this paper. 
We also propose a new method to determine the average GRB spectrum
in which the curvature of GRB spectrum compared to a pure power-law
is taken into account based on the hardness-brightness correlation
seen in the BATSE catalog.

The paper is organized as follows: in \S 2, formulations used in this
paper are described. We determine
the average spectrum of the short and long GRBs and give the relations
between redshift and peak flux of GRBs.  In \S 3, the models
of cosmic star formation history used in this paper are described.
The results of fits to the observed GRB distribution of the 4B BATSE
catalog (Paciesas et al. 1997) 
are given in \S 4. After discussing the result and its implications on
the cosmic star formation history in \S 5, 
we conclude this paper in \S 6. Unless otherwise mentioned, we use the
Einstein-de Sitter (EdS) universe with $(h, \Omega_0, \Omega_\Lambda)$
= (0.5, 1, 0) as background cosmology.

\section{Formulations}
Expected number of GRBs in a flux interval between $P_1$ and $P_2$
($P_1 < P_2$),
where $P$ is peak photon flux in cm$^{-2}$ sec$^{-1}$, is given by
\begin{equation}
N(P_1, P_2) = \zeta T_{\rm obs} \int_{z(P_2)}^{z(P_1)} \frac{dV}{dz} 
\frac{R_{\rm GRB}(z)}{(1+z)} dz \ ,
\label{eq:dNdP}
\end{equation}
where $dV/dz$ is the comoving volume element per unit redshift,
$\zeta$ the sky coverage efficiency [$\sim$ 38 \% for the BATSE experiment
(Meegan et al. 1996)], $T_{\rm obs}$ the observation term,
and $R_{\rm GRB}$ the comoving GRB rate density.
The factor $(1+z)^{-1}$ is included to account for the time dilation of 
the interval between detected bursts. We use the standard candle 
approximation throughout this paper and hence redshift and 
peak flux of GRB are related uniquely as $z(P)$. The comoving volume
element can be written as ($c=1$):
\begin{equation}
\frac{dV}{dz} = \frac{4 \pi d_L^2}{(1+z)} \left| \frac{dt}{dz}\right| \ ,
\end{equation}
where $t$ is the cosmic time and $dt/dz$ in the Freedman universe is given as
\begin{equation}
\left( \frac{dt}{dz} \right)^{-1} = - H_0 (1+z) \sqrt{
(1 + \Omega_0 z)(1+z)^2 - \Omega_\Lambda (2z + z^2) } \ ,
\end{equation}
and the luminosity distance $d_L$ is given as 
\begin{equation}
d_L = \left\{
\begin{array}{ll} 
\frac{1+z}{H_0 \sqrt{\Omega_0 + \Omega_\Lambda - 1}} 
\sin(d_{\rm prop} H_0 \sqrt{\Omega_0 + \Omega_\Lambda - 1} ) \ ,
& (\Omega_0 + \Omega_\Lambda > 1) \\
(1+z) d_{\rm prop} \ , & (\Omega_0 + \Omega_\Lambda = 1) \\
\frac{1+z}{H_0 \sqrt{1 - \Omega_0 - \Omega_\Lambda}} 
\sinh(d_{\rm prop} H_0 \sqrt{1 - \Omega_0 - \Omega_\Lambda} ) \ ,
& (\Omega_0 + \Omega_\Lambda < 1)  \ .
\end{array}  \right.
\end{equation}
Here the proper distance $d_{\rm prop}$ is given by
\begin{equation}
d_{\rm prop} = \int_0^z (1+z) \left| \frac{dt}{dz} \right| dz \ .
\end{equation}

The most important fitting parameter in the GRB brightness distribution
analysis is the distance scale of GRBs, and in this work we choose $\zmax$, 
the redshift of the most distant bursts corresponding to a threshold flux
of the analysis, $P_{\rm min}$. 
The $z$-$P$ relation is determined by the following equation:
\begin{equation}
\frac{P}{P_{\rm min}} = \frac{1+z}{1+\zmax} \frac{d_L^2(\zmax)}{d_L^2(z)}
\frac{k_P(z)}{k_P(\zmax)} \ ,
\label{eq:z-P}
\end{equation}
where $k_P(z)$ is the $k$-correction for the GRB photon spectrum:
\begin{equation}
k_P(z) = \frac{\int_{(1+z) \eps_l}^{(1+z) \eps_u} \frac{dL_n}{d\eps} d\eps}
{\int_{\eps_l}^{\eps_u} \frac{dL_n}{d\eps} d\eps} \ .
\end{equation}
Here $dL_n/d\eps$ is the differential photon number luminosity
of GRBs at rest, and $\eps$ the photon energy. The energy range
in which the peak flux is measured, ($\eps_l, \eps_u$), is 50--300 keV
for the BATSE experiment. 
The factor $(1+z)/(1+\zmax)$ in Eq. \ref{eq:z-P} is introduced because
the standard definition of $d_L$ is valid for energy flux but $P$ is
photon number flux.  The BATSE peak flux is given in three
different time scales of 64, 256, and 1024 msec. We use the 1024 msec
time scale for the long GRBs because the sensitivity is best in this 
scale, but 64 msec for the short GRBs because the average duration
of the short GRBs is $\sim$ 0.3 sec and we have to use shorter time scale
than this for the real peak flux. We set the threshold values of
$P_{\rm min}$ = 0.4 cm$^{-2}$sec$^{-1}$ (1024 msec) and
1.6 cm$^{-2}$sec$^{-1}$ (64 msec) for the long and short GRBs, respectively,
above which the BATSE trigger efficiency is better than 90 \%
(Paciesas et al. 1997). There are 773
long GRBs and 192 short GRBs above these thresholds in the 4B catalog.

\begin{figure}
  \begin{center}
    \leavevmode\psfig{figure=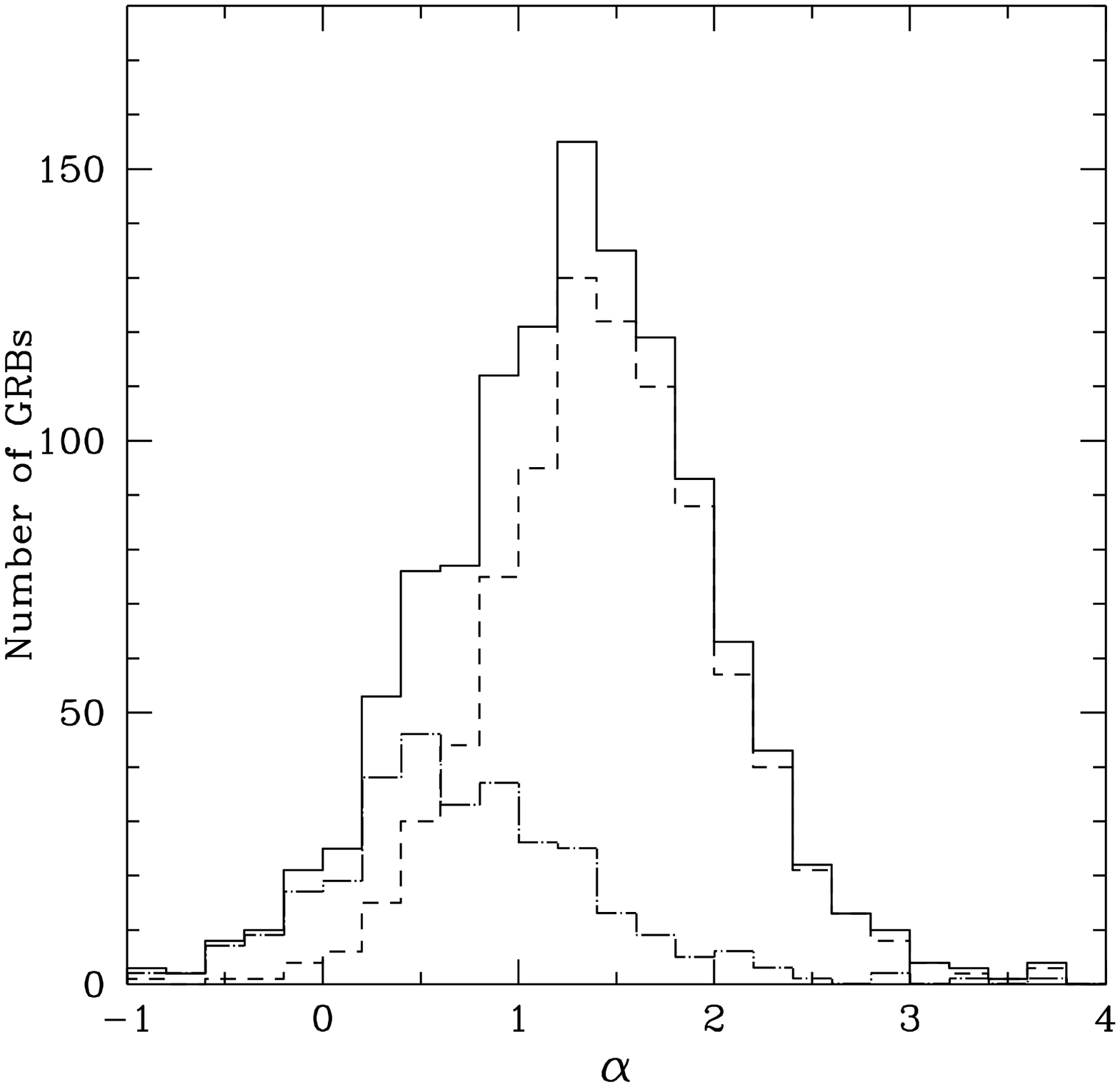,width=10cm}
  \end{center}
\caption{The histogram of effective photon index of GRB spectra ($\alpha$)
in the 4B BATSE catalog (Paciesas et al. 1997). 
The photon index is calculated from the
fluence ratio of the 50--100 keV and 100--300 keV ranges. The dashed line
is for the long GRBs, the dot-dashed line for short GRBs, and the solid
line for the total.}
\label{fig:alpha-hist}
\end{figure}

Now let us determine the photon spectrum of GRBs. Because the spectrum
is significantly different for the long- and short-duration bursts, 
we have to determine the spectra of these populations separately.
Figure \ref{fig:alpha-hist} shows a histogram of power-index of
photon spectra $(dL_n/d\eps \propto \eps^{-\alpha})$
for the long GRBs (dashed line), short GRBs (dot-dashed line),
and total (solid line). The values of $\alpha$ are estimated from
the ratio of the fluence in 50--100 keV and 100--300 keV in the
4B catalog. 
The average of $\alpha$ for long+short GRBs is 1.38, which is
consistent with a recent estimate of $\alpha = 1.1 \pm 0.3$
(Mallozzi, Pendleton, \& Paciesas 1996). 
It is clear from this figure that the long-duration
bursts have significantly softer spectra than short bursts;
average $\alpha$ is 1.52 and 0.82 for the long and short GRBs, respectively.

\begin{figure}
  \begin{center}
    \leavevmode\psfig{figure=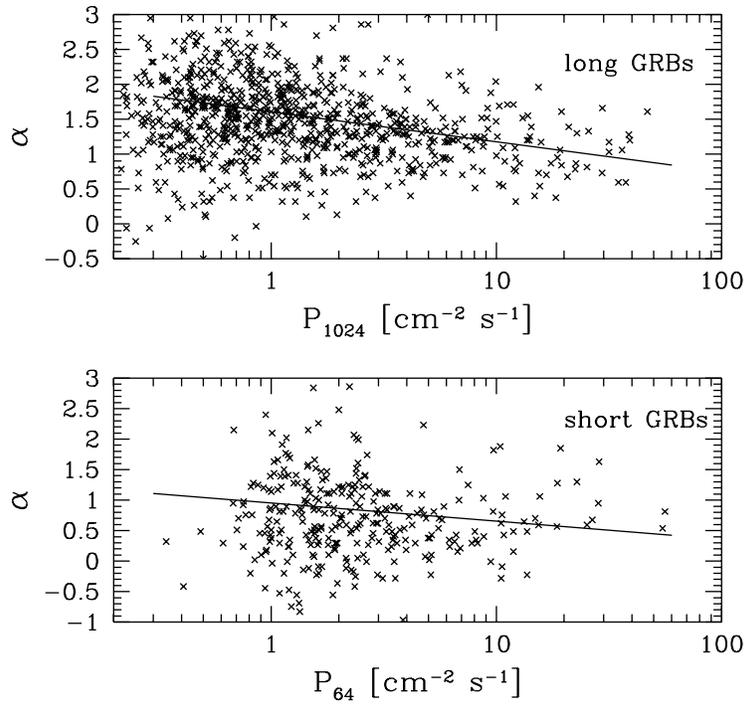,width=10cm}
  \end{center}
\caption{Correlation between the photon index ($\alpha$) and 
peak flux (1024 msec for the long GRBs and 64 msec for the short GRBs),
in the 4B BATSE catalog (Paciesas et al. 1997). 
The solid lines are least-square fits.}
\label{fig:alpha_P}
\end{figure}

\begin{figure}
  \begin{center}
    \leavevmode\psfig{figure=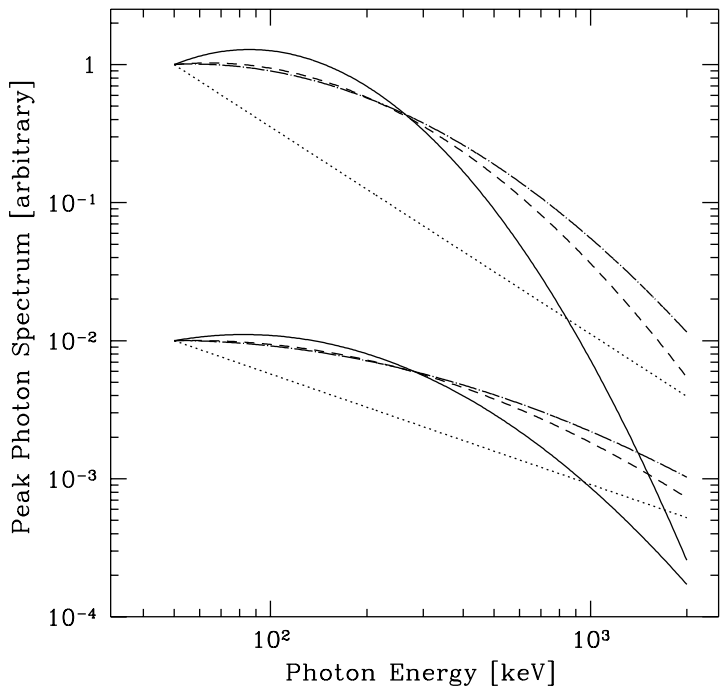,width=10cm}
  \end{center}
\caption{The photon spectrum of GRBs used in this paper. In our treatment,
the spectrum is determined when the fitting parameter, $\zmax$ is fixed
(see text).
The curvature reflects the $\alpha$-$P$ correlation shown in Fig. 
\protect\ref{fig:alpha_P}. The upper three curves correspond to the 
long GRBs and
the lower to the short GRBs. The solid, dashed, and dot-dashed lines
are for the cases of $\zmax$ = 1, 3, and 5, respectively. The dotted
lines are simple power-law spectra with the average photon indices
of the 4B catalog ($\alpha$ = 1.52 and 0.82 for the long and short GRBs,
respectively).
}
\label{fig:spec}
\end{figure}

It is an easy option to analyze the GRB distribution with these average
photon indices, but here we take a further step to include
the effect of curvature of GRB spectra. Figure \ref{fig:alpha_P}
shows $\alpha$-$P$ plots for the long and short GRBs. Both populations
(especially the long GRBs) show a trend that weaker bursts have softer
spectra, and this can be understood as follows. GRB spectra cannot be
described by a
pure power-law but exhibit some curvature, i.e., the spectrum becomes
softer with increasing photon energy (e.g., Mallozzi, Pendleton, \& Paciesas
1996). Because weaker bursts are more distant and hence more redshifted,
the observed spectrum becomes softer for weaker GRBs on average. 
In the following
we describe the determination of the GRB spectrum which takes account
of this $\alpha$-$P$ correlation in a consistent way.
First we assume a unique relation between $\alpha$ and $P$,
which is obtained by a linear fit for the observed $\alpha$-$P$ relation,
as shown in Fig. \ref{fig:alpha_P} by the solid lines.
We set a maximum peak flux $P_{\rm max}$, for which we adopt
60 cm$^{-2}$ sec$^{-1}$ both for the long and short GRBs, and denote the
photon indices corresponding to $P_{\rm min}$ and $P_{\rm max}$
as $\alpha_{\rm Pmin}$ and $\alpha_{\rm Pmax}$, respectively.
We also assume a functional form of $dL_n/d\eps$:
\begin{equation}
\log \frac{dL_n}{d\eps} = \frac{1}{2} A (\log \eps )^2 + B \log \eps
+ {\rm const.} 
\end{equation}
Although not physically meaningful, this form describes the
four-channel burst spectra of the BATSE GRBs
quite well (Mallozzi, Pendleton, \& Paciesas 1996).
The two parameters, $A$ and $B$, are determined so that the average
photon index in the restframe energy range of $[(1+z)\varepsilon_l,
(1+z)\varepsilon_u]$ is consistent with the
$\alpha$-$P$ relation:
\begin{eqnarray}
A \langle \log \eps\rangle_{P_{\rm max}} + B &=& - \alpha_{\rm Pmax} \\
A \langle\log \eps\rangle_{P_{\rm min}} + B &=& - \alpha_{\rm Pmin} \ ,
\end{eqnarray}
where
\begin{eqnarray}
\langle \log \eps\rangle_{P_{\rm max}} &=& \frac{1}{2}
\{ \log[(1+z_{\rm min})\eps_l] + \log[(1+z_{\rm min})\eps_u] \} \\
\langle \log \eps\rangle_{P_{\rm min}} &=& \frac{1}{2}
\{ \log[(1+z_{\rm max})\eps_l] + \log[(1+z_{\rm max})\eps_u] \} \ .
\end{eqnarray}
The above spectrum is determined if $\zmax$ and $z_{\rm min}$, which are
the redshifts corresponding to $P_{\rm min}$ and $P_{\rm max}$, are
fixed. The fitting parameter $\zmax$ can be chosen arbitrary and
the unknown parameter $z_{\rm min}$
is determined by solving the following consistency equation:
\begin{equation}
\frac{P_{\rm max}}{P_{\rm min}} = 
\frac{1+z_{\rm min}}{1+\zmax} \frac{d_L^2(\zmax)}{d_L^2(z_{\rm min})}
\frac{k_P(z_{\rm min})}{k_P(\zmax)} \ .
\end{equation}
Now the GRB spectrum, which depends on the fitting parameter $\zmax$,
has been determined. We note that the above treatment of the GRB spectrum is
the first to take account of the $\alpha$-$P$ correlation in a consistent
way. In Figure 
\ref{fig:spec}, we show some examples of the GRB spectrum determined
by the above method. The spectral curvature becomes larger with
decreasing $\zmax$ because the restframe spectrum has to become
softer more rapidly with increasing photon energy.

\begin{figure}
  \begin{center}
    \leavevmode\psfig{figure=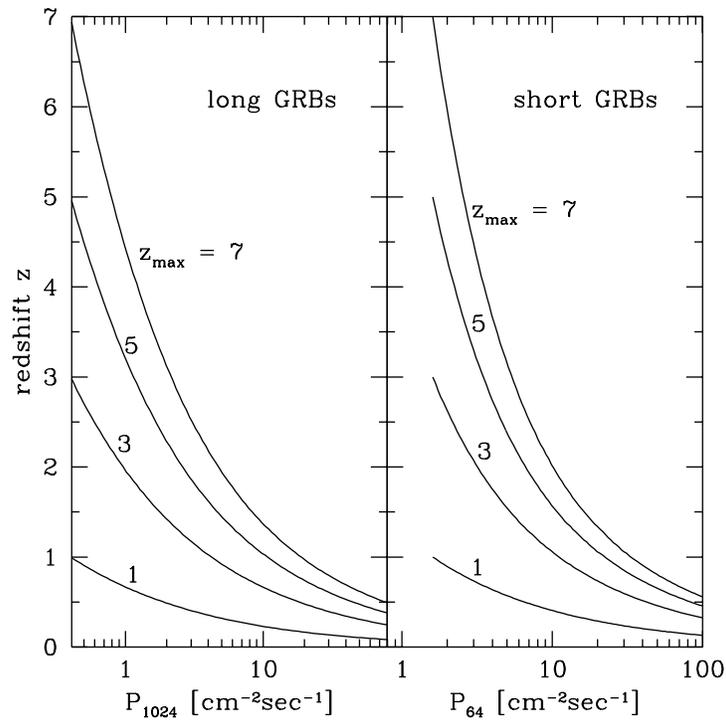,width=10cm}
  \end{center}
\caption{The redshift vs. peak flux relation for the long and short GRBs,
for some values of $\zmax$ which is the redshift corresponding to
$P_{1024} = 0.4$ and $P_{64} = 1.6 \ \rm cm^{-2} sec^{-1}$ for the 
long and short GRBs. The values are indicated in the figure. 
}
\label{fig:z-P}
\end{figure}

\begin{figure}
  \begin{center}
    \leavevmode\psfig{figure=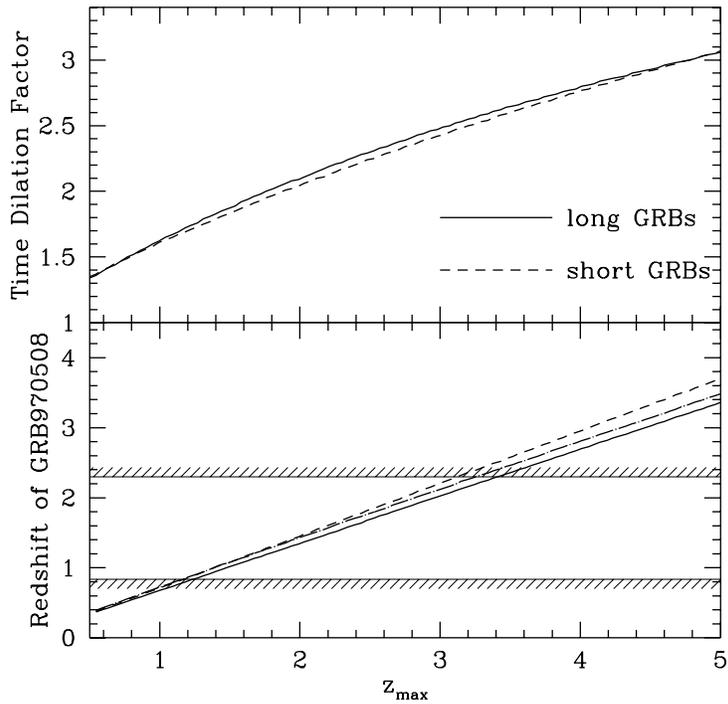,width=10cm}
  \end{center}
\caption{Time dilation factor
and redshift of GRB 970508, as functions of $\zmax$. For the definition of
time dilation factor, see text. In the upper panel, the solid line is for
the long GRBs and the dashed line for the 
short GRBs. In the lower panel, the three
lines are all for the long GRBs with different cosmological parameters:
$(h, \Omega_0, \Omega_\Lambda) = (0.5, 1, 0), (0.6, 0.2, 0),$ and
(0.7, 0.2, 0.8) for the solid, dashed, and dot-dashed lines, respectively.
The constraint of $0.835 \leq z < 2.3$ which is
set by the absorption lines (Metzger et al. 1997) is also shown.
}
\label{fig:td}
\end{figure}

Now the relation between $z$ and $P$ is fixed, and the $z$-$P$ plots
are shown in Fig. \ref{fig:z-P} both for the long and short GRBs
for some values of $\zmax$. 
Since the $z$-$P$ relation has been fixed, we can calculate
the cosmological time dilation factor and redshift of some
particular GRBs as functions of $\zmax$. Cosmological time dilation
effect has been searched for various temporal structures, and a
dilation factor of about 2 for the long duration bursts
is widely accepted (see, e.g., Norris et al.
1994, 1995; Fenimore \& Bloom 1995). Recently Che et al. (1997)
suggested the existence of time dilation of about 2
also for the short GRBs.
In the upper panel of Fig. \ref{fig:td}, we plot the time dilation
factor, $(1+z_{\rm weak})/(1+z_{\rm bright})$,
as a function of $\zmax$. The brighter and weaker
peak fluxes are taken as
$(P_{\rm weak}, P_{\rm bright})$ = (0.46, 9.5) and (1.6, 10.0)
for the long and short GRBs, respectively. The values for the long GRBs
correspond to the ``dim+dimmest'' and  ``bright'' populations
defined by Norris et al. (1994, see also Horack, Mallozzi, \& Koshut 1996
for estimate of peak flux for these groups), 
and those for the short GRBs
roughly correspond to the minimum and maximum of the brightness 
parameter defined by Che et al. (1997). 
The dilation factor of about 2
suggests $\zmax \sim$ 2 both for the long and short GRBs. The lower panel of 
this figure shows the redshift of GRB970508, which is a long GRB
with 1024 msec peak flux of 0.85 $\rm cm^{-2} sec^{-1}$ (when 
$\alpha = 1.5$, see Kouveliotou et al. 1997),
as a function of $\zmax$. The constraint of $0.835 \leq z < 2.3$
for this GRB (Metzger et al. 1997) suggests $1.22 \leq \zmax < 3.43$,
although possible dispersion in GRB luminosity might relax this
constraint.

Although we have chosen the maximum redshift as 
the distance parameter of GRBs, 
it would be convenient to give relations between
$\zmax$ and absolute luminosity or total emitted energy of GRBs.
If the gamma-ray emission is isotropic,
the absolute peak luminosity in the restframe energy range of $(\eps_l,
\eps_u$) is given as:
\begin{equation}
L_{\rm peak} = \langle \eps \rangle \frac{4 \pi d_L(\zmax)^2 P_{\min}}
{k_P(\zmax) (1+\zmax)} \ ,
\end{equation}
where $\langle \eps \rangle$ is the average energy 
in restframe energy range of $(\eps_l, \eps_u$).
The total energy emitted by GRBs in the same energy range
can be estimated from $f_{\min}$,
which is the fluence of GRBs at $z = \zmax$:
\begin{equation}
E_{\rm tot} =  \frac{4 \pi d_L(\zmax)^2 f_{\min}}
{k_e(\zmax) (1+\zmax)} \ ,
\end{equation}
where $k_e(z)$ is the $k$-correction for the energy spectrum,
\begin{equation}
k_e(z) = \frac{\int_{(1+z) \eps_l}^{(1+z) \eps_u} \eps 
\frac{dL_n}{d\eps} d\eps}
{\int_{\eps_l}^{\eps_u} \eps \frac{dL_n}{d\eps} d\eps} \ .
\end{equation}
We have estimated $f_{\min}$ from fluence-$P$ plots for the BATSE catalog.
The peak luminosity and total energy emitted
in the restframe energy range of 50--300 keV
are shown in Fig. \ref{fig:luminosity} as functions of $\zmax$, as well as 
the values after bolometric correction (50--2000 keV).
Since the spectrum becomes harder,
the bolometric correction factor becomes larger with increasing $\zmax$.
It should be noted that the bolometric correction is based on a simple
extrapolation of
the assumed functional form of $dL_n/d\varepsilon$, but there might
be cut-off energy in the GRB spectrum in reality. Therefore the estimate of
bolometric luminosity or total energy has to be taken with care, especially
for the short GRBs.

\section{Modeling the Cosmic GRB Rate History}
\begin{figure}
  \begin{center}
    \leavevmode\psfig{figure=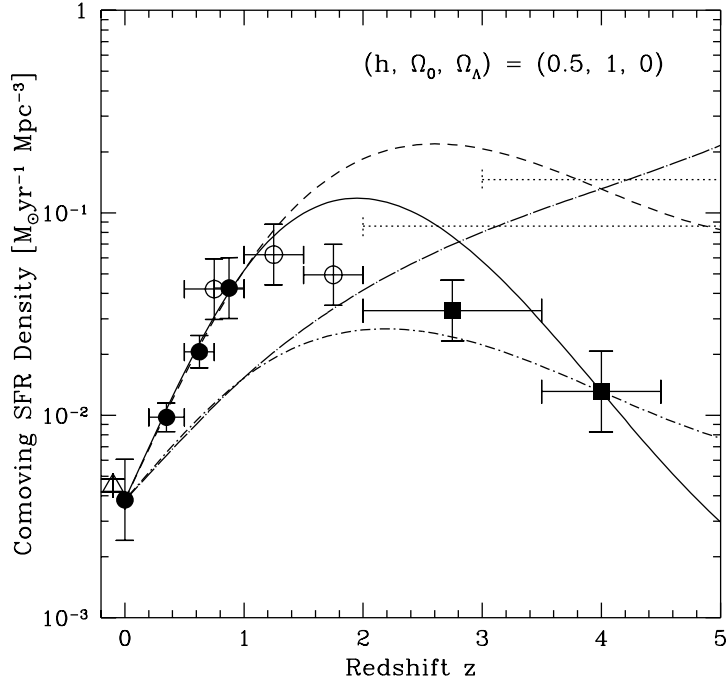,width=10cm}
  \end{center}
\caption{Cosmic star formation history as a function of redshift. Data points
are direct observational estimates by Gallego et al. (1995, open triangle),
Lilly et al. (1996, filled circles), Connolly et al. (1997, open circles),
and Madau et al. (1997, filled square).  Curves are the model of cosmic
star formation history used in this paper, with the parameters
($\xi(1), \xi(4))$ = (13.6, 3.43), (13.6, 34.3), (4, 3.43), and (4, 34.3)
for the 
solid, dashed, short-dot-dashed, and long-dot-dashed lines, respectively
(see text in detail).
The two horizontal dotted lines are average star formation rates expected
if the stars in present-day elliptical galaxies were formed before
$z$ = 2 or 3. }
\label{fig:csfh}
\end{figure}

We can now calculate the GRB brightness distribution and compare
it to the observation, provided that $R_{\rm GRB}$ is determined
as function of $z$. In this paper we investigate two possibilities:
the proportional model and the NS-NS model.
Figure \ref{fig:csfh} shows observational estimations of comoving
SFR density based on H$\alpha$ or
UV luminosity density (Gallego et al. 1995; Lilly et al. 1996; Connolly 
et al. 1997; Madau, Pozzetti, \& Dickinson 1998). 
In the Einstein-de Sitter
universe with $(\Omega_0, \Omega_\Lambda) = (1, 0)$, the cosmic SFR
at $z=1$ is about 15 times higher than the local value, and
the data at higher redshifts suggest a peak of cosmic SFR at 
$z \sim$ 1--2. Let $\xi(z)$ be the ratio of SFR at $z$ to the local SFR,
i.e., $\xi (z) = R_*(z)/R_*(0)$, where $R_*$ is the comoving 
SFR density in the universe in [$M_\odot$yr$^{-1}$Mpc$^{-3}$].
In this paper, we parametrize the cosmic star formation
history by two parameters of $\xi(1)$ and $\xi(4)$, and SFR evolution
is determined by the natural cubic spline
interpolation (e.g., Press et al. 1992)
between SFRs at $z = 0, 1,$ and 4. From the observational data, we adopt
$\xi(1) = 13.6$ and $\xi(4) = 3.43$ as the standard observational
star formation history (solid line in Fig. \ref{fig:csfh}).

Although the UV luminosity density evolution in $z$ = 0--1
measured by the Canada-France
Redshift Survey (CFRS; Lilly et al. 1996) is very steep, another 
estimate of this evolution favors more modest evolution. Totani, 
Yoshii, \& Sato (1997; hereafter TYS) 
pointed out that the present-day colors of galaxies
are strongly correlated to recent star formation history in them
and it is possible to estimate the SFR evolution at $z \ltilde 1$ from
stellar population synthesis models of galaxies. TYS summed up the star 
formation history of galaxy evolution models for five types of galaxies, 
E/S0, Sab, Sbc, Scd, and Sdm, according to observed relative proportion of
these types.  Each of the five models is made to reproduce the
spectral energy distribution in a wide range of wavelength at present and
observed chemical properties 
(Arimoto \& Yoshii 1987; Arimoto, Yoshii, \& Takahara 1992). 
The estimate of TYS 
for $\xi(1)$ is about 4 if the age of the universe is reasonable
(10--15 Gyr). This value is significantly lower than the result of
Lilly et al. (1996), but if one assumes a $\Lambda$-dominated universe
with $(\Omega_0, \Omega_\Lambda$) = (0.2, 0.8) the UV-estimated $\xi(1)$
becomes smaller by a factor of 1.8 and the discrepancy is a little remedied.
Furthermore, 
Hammer et al. (1997) pointed out another effect which may be responsible
for the very high $\xi(1)$ in the CFRS; star forming galaxies in the CFRS
at $z > 0.5$ seem to have significantly low metallicities and 
hence they may be more transparent to UV light than local galaxies.
This effect makes the evolution of UV luminosity density steeper than
SFR evolution, and SFR evolution based on UV
observations is overestimated when this effect is neglected.
We will later discuss these points after we obtain the constraints
on $\xi(1)$ from the GRB distribution.

It is also widely discussed that the UV-estimated SFR 
at $z>1$ is highly uncertain
and may be significantly underestimated because of the dust extinction 
effect (Meurer et al. 1997; Pettini et al. 1997; Sawicki \& Yee 1998;
Heckman 1998; Cimatti et al. 1998). 
The ratio of far infrared luminosity to UV in
metal-rich ($\sim Z_\odot$) local starburst galaxies is 10--100 
(Heckman 1998) and
this suggests that high redshift galaxies at $z>2$, which also
show signatures of starbursts, might be significantly affected by 
interstellar dust extinction by factor of $\sim 10$
(Meurer et al. 1997; Sawicki \& Yee 1998) or in some cases $\sim 500$
(Cimatti et al. 1998).  This possible upward
correction for high redshift SFR data is interesting because 
the formation of spheroidal systems such as elliptical galaxies
or bulges of spiral galaxies also predicts higher SFR at $z \gtilde 2$, 
if they were formed by starbursts at high redshifts.
In order to give a quantitative estimate, let us calculate the
total amount of star formation in elliptical galaxies. The total
B-band luminosity density at present 
is $1.2 \times 10^8 h L_{B \odot}$ Mpc$^{-3}$ (Lilly et al. 1996),
where $h = H_0/ \rm (100km \ s^{-1}Mpc^{-1})$, 
with the fraction of elliptical galaxies
of 0.28 (Totani, Yoshii, \& Sato 1997).
By using the mass-to-luminosity
ratio $(M/L)_B = 16.2 h$ for elliptical galaxies (Faber \& Gallagher 1979),
the total amount of stars which have been formed in elliptical galaxies
becomes 7.6 $\times 10^8 h^2$ [$M_\odot$ Mpc$^{-3}$]. Here we have taken
account of a recycling factor of 1.4 inferred from a galaxy evolution
model of Arimoto \& Yoshii (1987). The horizontal dotted 
lines in Fig. \ref{fig:csfh}
are the average SFR expected in elliptical galaxies if they
were formed before $z=2$ or 3. This result suggests that
the upward correction by a factor of $\sim$ 10 is very likely
from the viewpoint of spheroidal system formation.
Considering the uncertainties discussed above,
we will try various values of $\xi(1)$ and $\xi(4)$
in the following GRB analysis. 
We have plotted some examples of the model SFR evolution with different values
of $\xi(1)$ (=13.6 or 4) and $\xi(4)$ (=3.43 or 34.3) 
in Fig. \ref{fig:csfh}.

\begin{figure}
  \begin{center}
    \leavevmode\psfig{figure=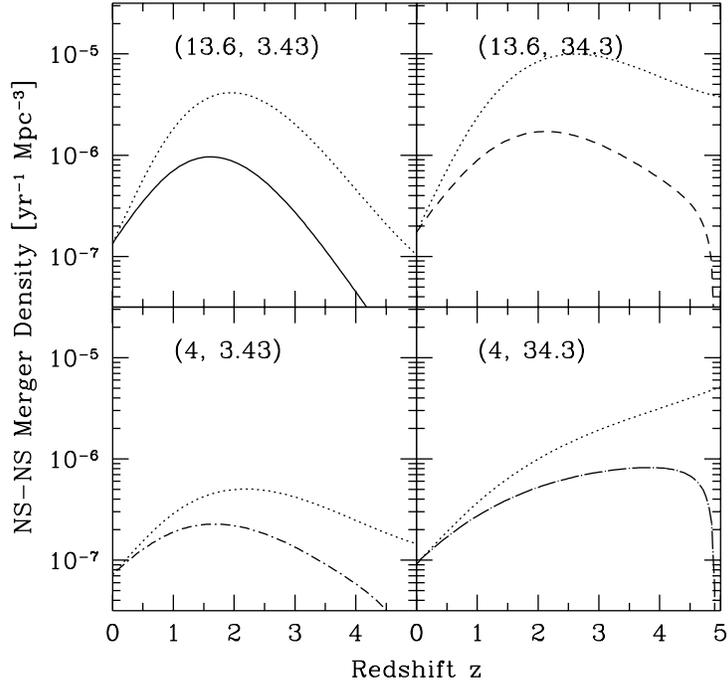,width=10cm}
  \end{center}
\caption{The merger rate evolution of binary neutron stars as a function of 
redshift, calculated from the cosmic SFR evolution shown in Fig. 
\protect\ref{fig:csfh}. The four panels correspond to the different four
models of cosmic SFR evolution with values of $(\xi(1), \xi(4))$
indicated in the figure. The line marking corresponds to that of
Fig. \protect\ref{fig:csfh}. The dotted lines are the SFR
evolution before convolved with the time delay, 
normalized at $z=0$ for comparison. 
The time delay between star formation and merger makes the rate evolution 
flatter in $z$ = 0--1.}
\label{fig:ns2-rate}
\end{figure}

If GRBs are associated to NS-NS mergers, we cannot use the simple
approximation of $R_{\rm GRB} \propto R_*$, and 
the GRB rate evolution is written as:
\begin{equation}
R_{\rm GRB}(t) = \int_0^t P_m(t-t') R_*(t') dt' \ ,
\end{equation}
where $t$ is the cosmic time and $P_m(x)$ is the probability
distribution of time delay from star formation to NS-NS mergers
per unit mass of star formation,
in units of [$M_\odot^{-1}$ yr$^{-1}$].
The form of $P_m$ can be well approximated as $P_m(x) =
A (x/t_c)^\gamma$ with $t_c \sim$ 0.02 Gyr and $\gamma \sim -1$
(T97). More detailed calculations of $P_m$ by binary population
synthesis suggest the uncertainty in $\gamma$ is probably about 0.2--0.3
(Lipunov et al. 1997; Portegies Zwart and Yungelson 1998).
Because the aim of this paper is to investigate the cosmic star 
formation history from GRBs, we fix the parameters $t_c$ and $\gamma$
to 0.02 Gyr and $-1$, respectively. The normalization factor, $A$,
can be inferred also from binary population synthesis calculations.
Lipunov et al. (1995) estimated the present merger rate as
2 $\times 10^{-4}$ [yr$^{-1}$], when $10^{11} M_\odot$ stars have been
formed constantly. Portegies Zwart \& Yungelson (1998), by similar
calculations, found the present rate as 3.4 $\times 10^{-5}$ yr$^{-1}$,
when the SFR is constant at 4 $M_\odot$yr$^{-1}$ and the present age
of the Galaxy is 10 Gyr. Requiring these values to be consistent
with the form of $P_m$, we find $A$ = 1.6 $\times 10^{-13}$
and 7 $\times 10^{-14}$ [$M_\odot^{-1}$ yr$^{-1}$] for the 
results of the two groups, respectively. We use $A = 1 \times 10^{-13}$ 
[$M_\odot^{-1}$ yr$^{-1}$] for
calculation of the absolute rate of NS-NS mergers and the uncertainty in $A$ is
probably about a factor of 2, as suggested from the difference of
the two groups. This uncertainty affects the overall normalization of 
GRB rate and hence estimate of the beaming factor of GRBs, but does not affect
the evolution of GRB rate which is the main topic of this paper. In Fig. 
\ref{fig:ns2-rate} we show some examples of calculation of $R_{\rm GRB}(z)$,
generated from the four models of $R_*(z)$ shown in Fig. \ref{fig:csfh}.
The GRB rate evolution becomes flatter than the star formation history
when we take account of the time delay, and 
the importance of this effect is clear if 
one compares these results to those neglecting the delay (dotted lines).

\section{Comparison to the BATSE Data}
\subsection{Brightness Distribution of GRBs}
T97 used the Kolmogorov-Smirnov test for the comparison between
the observation and the model calculation, which utilizes the
cumulative flux distribution, $N(>P)$. However, the differential
flux distribution, $dN/dP$, reflects the GRB rate evolution better
than $N(>P)$ and we use the $\chi^2$ analysis for the differential
distribution in this paper.
We set 9 bins in $(P_{\min}, P_{\max})$ with a logarithmically
constant interval
both for the long and short bursts. The expected number of GRBs in each
bin is calculated by using eq. (\ref{eq:dNdP}).

\begin{figure}
  \begin{center}
    \leavevmode\psfig{figure=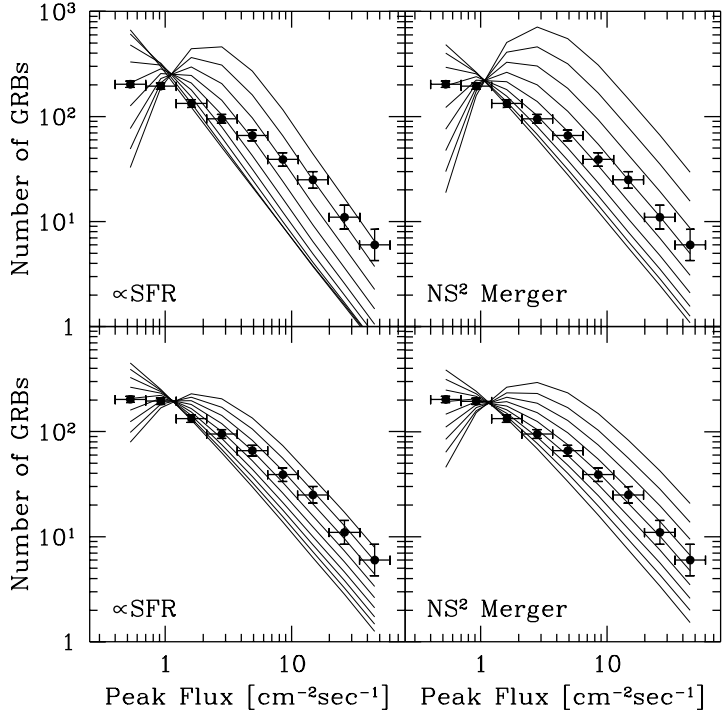,width=10cm}
  \end{center}
\caption{Differential distribution of peak flux (1024 msec) of the long GRBs.
The data points are the BATSE data and the error bars in peak flux
indicate the width of bins. Nine model curves are shown for each panel, 
corresponding to 
different values of $\zmax$ = 1--5 with an interval of 0.5. The number of GRBs
at small peak flux decreases with increasing $\zmax$ because of the low SFR
at high redshifts. The left panels are for the proportional model,
while the right panels for the NS-NS model. In the
upper panels we use $\xi(1) = 13.6$ while in the lower panels $\xi(1) =4$.
In all panels $\xi(4) = 3.43$.}
\label{fig:long1024}
\end{figure}

\begin{figure}
  \begin{center}
    \leavevmode\psfig{figure=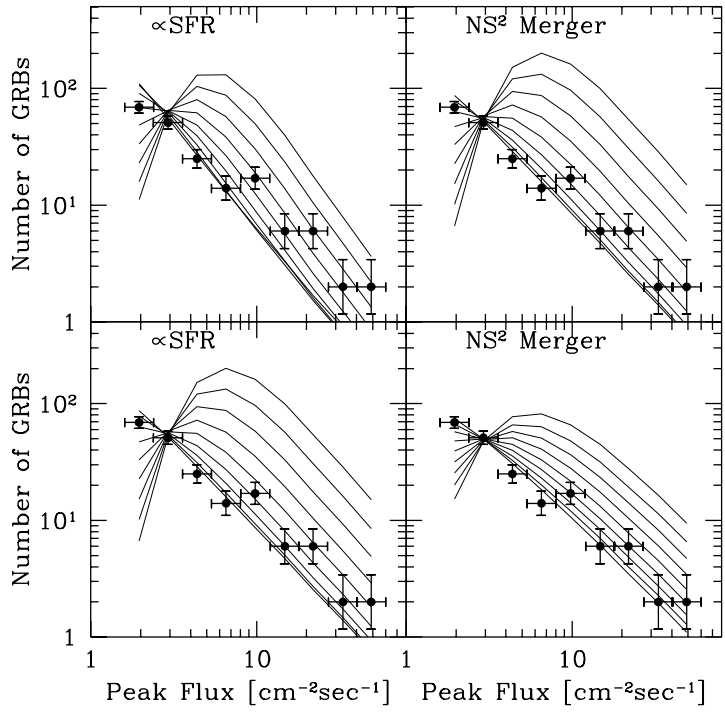,width=10cm}
  \end{center}
\caption{The same as Fig. \protect\ref{fig:long1024}, but for the short
GRBs with 64 msec peak flux.
}
\label{fig:short64}
\end{figure}

Adopting the standard observational star 
formation history, i.e., $\xi(1)=13.6$ and $\xi(4)=3.43$, we 
have calculated the expected number of GRBs in each bin with nine
values of $\zmax$ (1--5 with an interval of 0.5), 
and the results are shown in the upper panels of
Fig. \ref{fig:long1024} for the long bursts. (We have chosen the best-fit
normalizations for the model calculations and hence only the
shape of GRB flux distribution is tested.)
In the upper-left panel the GRB rate is assumed to be proportional to
SFR while the upper-right panel assumes the NS-NS merger scenario.
These panels show that the inclination of the model curves
is hardly changed by $\zmax$ in peak flux larger than $\sim$
10 cm$^{-2}$sec$^{-1}$, while in small $P$ range the expected event number 
becomes significantly smaller with increasing $\zmax$, because of 
the paucity of SFR at high $z$ in the observational star formation history.
The standard observational SFR evolution does not agree with the GRB 
distribution, in either the proportional or NS-NS model.
Because the inclination in the bright range of $P$ is almost insensitive
to $\zmax$, we can set a stringent constraint on the SFR evolution
at relatively low redshift ($z \ltilde 1$). Actually these results 
suggest that the SFR evolution inferred from the UV luminosity density
at $z<1$ (Lilly et al. 1996) is too steep to be consistent with
the GRB brightness distribution, both in the proportional model and
the NS-NS model. One way to get
a better fit is to decrease the value of $\xi(1)$.
In the lower panels, a lower value of $\xi(1)=4$, as inferred from the
galaxy evolution model (TYS), is used and then 
the discrepancy in the bright $P$ range is significantly remedied.
Figure \ref{fig:short64} is the same with Fig. \ref{fig:long1024}
but for the short GRB population. Although the statistics is poor, 
the UV-estimated SFR evolution in $z \sim$ 0--1 is too steep also 
for the short GRBs if the GRB rate is proportional to SFR.

\begin{figure}
  \begin{center}
    \leavevmode\psfig{figure=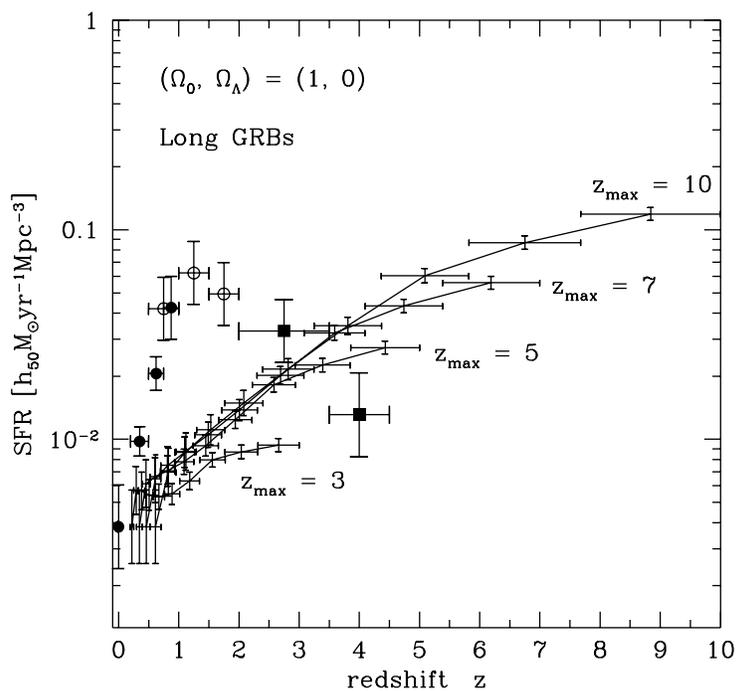,width=10cm}
  \end{center}
\caption{The star formation history in the universe deduced from the
brightness distribution of the long GRBs (crosses combined with solid lines). 
The GRB rate is assumed to be proportional to SFR. Four different values
of $\zmax$ are used, as indicated in the figure. The data points of
filled and open circles, and filled squares are the observational SFR estimates
as shown in Fig. \protect\ref{fig:csfh}.
}
\label{fig:SFR}
\end{figure}

The serious discrepancy between the UV-estimated SFR evolution
in $z$ = 0--1 and GRB distribution can also be seen in a SFR-$z$
diagram. When GRB rate is proportional to SFR, we can convert 
the observed number of GRBs directly into cosmic SFR. 
Let $N_{\rm obs}(P_1, P_2)$
be the observed number of GRBs in the interval of $P_1 < P < P_2$. Then
average SFR in the redshift interval of $z(P_2) < z < z(P_1)$ can be
written as:
\begin{equation}
R_* \propto \frac{N_{\rm obs}(P_1, P_2)}{\int_{z(P_2)}^{z(P_1)}
\frac{dV}{dz} \frac{dz}{(1+z)}} \ .
\end{equation}
Figure \ref{fig:SFR}
shows evolution of cosmic SFR calculated from the observed number of
long GRBs as a function of $z$ with various $\zmax$, where the
binning in $P$ is the same with that in Fig. \ref{fig:long1024}. 
These curves are normalized to the observed local SFR at minimum redshifts.
The SFR evolution inferred from GRBs
becomes steeper with increasing $\zmax$ but
the evolution in $z$ = 0--1 is much flatter than the UV-estimated evolution
even when we take an extreme value of $\zmax = 10$.
Therefore we conclude that the UV-estimated SFR evolution is
inconsistent with the hypothesis that brightness distribution of GRBs
traces the cosmic star formation history, if
the standard candle approximation is valid. In order to get an acceptable
fit, one must reduce the high SFR at $z=1$, and this is consistent with
the previous result of T97 who found that the star formation history
based on the galaxy evolution model of TYS (i.e., $\xi(1) \sim 4$) gives a
good fit but the observational SFR evolution does not. 
This result is also consistent
with that obtained by Petrosian \& Lloyd (1997), but does not agree
with that of Wijers et al. (1998). 

\subsection{Constraints on Cosmic SFR Evolution}
In this section we give quantitative constraints on the cosmic SFR evolution,
i.e., $\xi(1)$ and $\xi(4)$, by detailed statistical comparison of 
the theoretical and observed brightness distribution. 
In the $\chi^2$ test, there are four parameters both in the proportional
and NS-NS models: $\zmax, \xi(1), \xi(4)$, and normalization of the 
GRB rate. We have searched the best-fit values which give the minimum
chi-square ($\chi^2_{\min}$) in the two models of
GRB rate evolution both for the long and short GRBs, and the results are
listed in Table \ref{table:SFR-SFR}. The searched ranges are 
1--5 for $\zmax$ and 1--100 for $\xi(1)$ and $\xi(4)$.
The goodness-of-fit of the models
can be assessed by $\chi^2_{\min}$ with $9-4=5$ degrees of freedom.
The significance level (S.L.), 
i.e., the probability that $\chi_{\min}^2$ is larger than
the observed value by statistical fluctuation is also given in the table,
and large values of S.L. in the table show that the fitting is acceptable.

\begin{figure}
  \begin{center}
    \leavevmode\psfig{figure=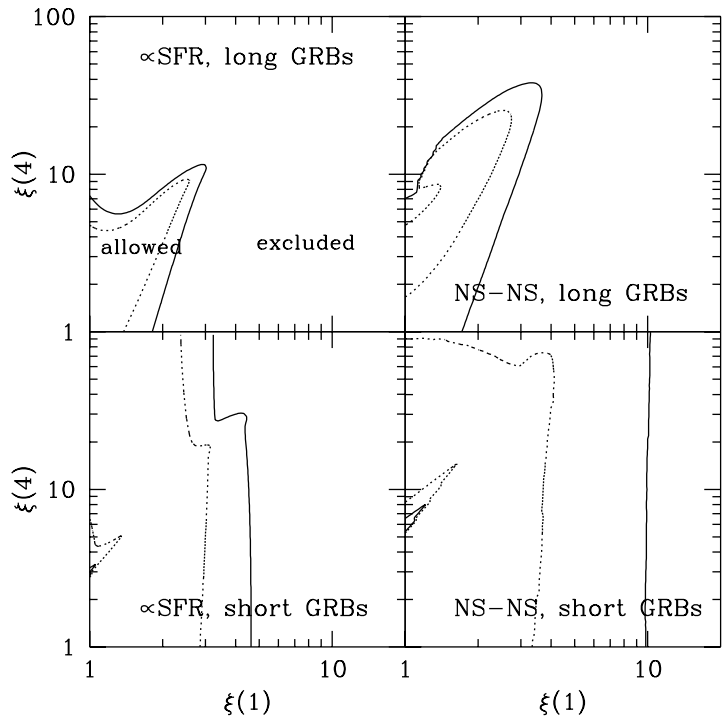,width=10cm}
  \end{center}
\caption{Constraints on the cosmic star formation history from GRB brightness
distribution. The allowed regions of $\xi(1)$ and $\xi(4)$ are shown
for the long and short GRBs (upper and lower panels, respectively) and 
for the proportional model and the NS-NS model (left and right panels, 
respectively). The solid line shows the 95 \% C.L. constraints, while the
dotted lines for 68 \% C.L. The Einstein-de Sitter universe with $h=0.5$
is assumed.}
\label{fig:SFR-SFR-EdS}
\end{figure}

\begin{figure}
  \begin{center}
    \leavevmode\psfig{figure=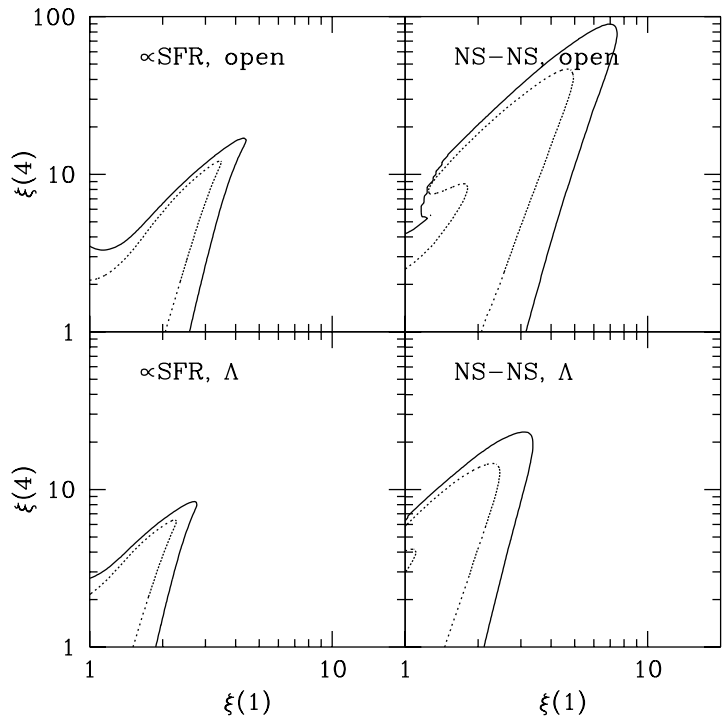,width=10cm}
  \end{center}
\caption{The same as Fig. \protect\ref{fig:SFR-SFR-EdS}, but only for the
long GRBs with different cosmological parameters: $(h, \Omega_0, 
\Omega_\Lambda)$ = (0.6, 0.2, 0) and (0.7, 0.2, 0.8) for the upper and
lower panels, respectively.
}
\label{fig:SFR-SFR-op-lam}
\end{figure}

Figure
\ref{fig:SFR-SFR-EdS} shows the allowed regions of $\xi(1)$ and $\xi(4)$
for the long and short GRBs, both for the proportional model
and the NS-NS model. Here the allowed region of the $\xi(1)$-$\xi(4)$ plane
is derived as follows. For each set of ($\xi(1)$, $\xi(4)$), we have searched
the best-fit value of $\zmax$ which gives a minimum value of chi-square,
which we denote as $\tilde\chi^2_{\min}[\xi(1), \xi(4)]$.
Then $\Delta \chi^2 \equiv \tilde\chi^2_{\min}[\xi(1), \xi(4)] -
\chi^2_{\min}$ is distributed as a chi-square distribution with 2 degrees of
freedom (Press et al. 1992). 
Therefore the 95\% C.L. region of ($\xi(1)$, $\xi(4)$) is given
by a contour of $\Delta \chi^2 = 5.99$. It should be noted that 
the constraints on ($\xi(1)$, $\xi(4)$) obtained as above are
conservative about the unknown value of $\zmax$.
As discussed in the previous section, the most
important result is the constraint on $\xi(1)$: for the long GRBs, 
$\xi(1) < 3.0$ (proportional) and $\xi(1) < 3.8$ (NS-NS) at 95 \% C.L.,
for any values of $\xi(4)$ and $\zmax$.
The time delay in the NS-NS scenario makes the GRB rate evolution 
flatter in $z$ = 0--1 and hence allows acceptable fits with larger 
$\xi(1)$ than the proportional model. The $\xi(1)$ is constrained
also for the short GRBs as $\xi(1) < 4.5$ and $\xi(1) < 10$ (95 \% C.L.),
for the proportional and NS-NS models, respectively.
If the SFR at $z=1$ is more than 10 times higher than the present rate
as suggested by the UV luminosity density, and if the standard candle
approximation is a good approximation,
the proportional model is completely rejected 
both for the long and short GRBs, and the NS-NS model is also
inconsistent with the distribution of the long GRBs.
In order to see the effect of changing the cosmological parameters,
we show the same contour maps in Fig. \ref{fig:SFR-SFR-op-lam}
only for the long GRBs in the two different
cosmological models: open $[(h, \Omega_0, \Omega_\Lambda) = (0.6, 0.2, 0)]$
and lambda $[(h, \Omega_0, \Omega_\Lambda) = (0.7, 0.2, 0.8)]$. The constraint
on $\xi(1)$ becomes weaker in the open universe, but the upper limit
for $\xi(1)$ in the proportional model is still much smaller than 10.

\begin{figure}
  \begin{center}
    \leavevmode\psfig{figure=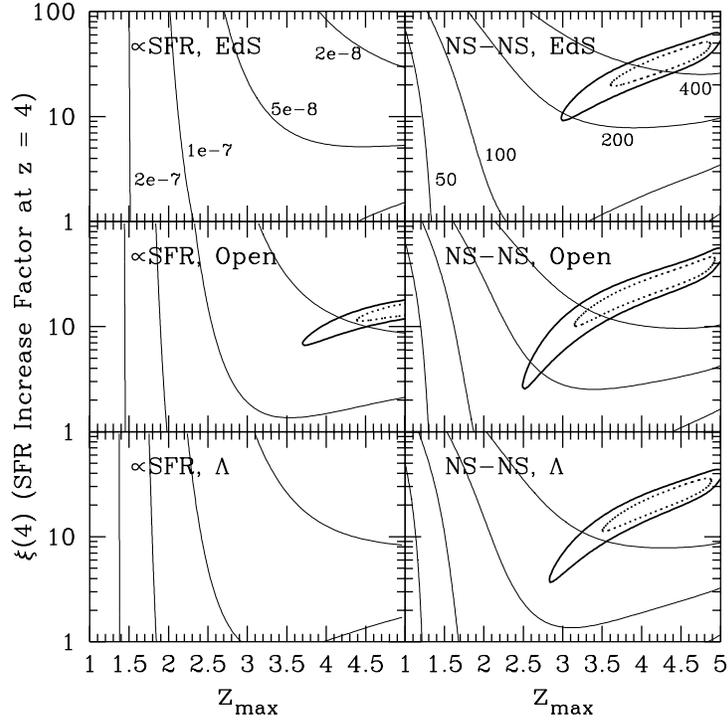,width=10cm}
  \end{center}
\caption{Allowed regions in $\zmax$-$\xi(4)$ plane are shown, assuming
$\xi(1) = 4$. The left panels are for the proportional model and 
the right for the NS-NS model, with the three different sets of cosmological
parameters for each, as shown in the figure. The thick-solid line
is 95 \% C.L. constraint and the thick-dotted line for 68 \% C.L. 
In the upper-left and lower-left panels, there is no allowed
region. The
thin solid lines are contours of the normalization factor: production
rate of GRBs per unit mass of star formation [GRB $M_\odot^{-1}$]
in the left panels and beaming factor ($4 \pi/\Delta \Omega)$ in the
right panels. 
}
\label{fig:zmax-SFR}
\end{figure}

\begin{figure}
  \begin{center}
    \leavevmode\psfig{figure=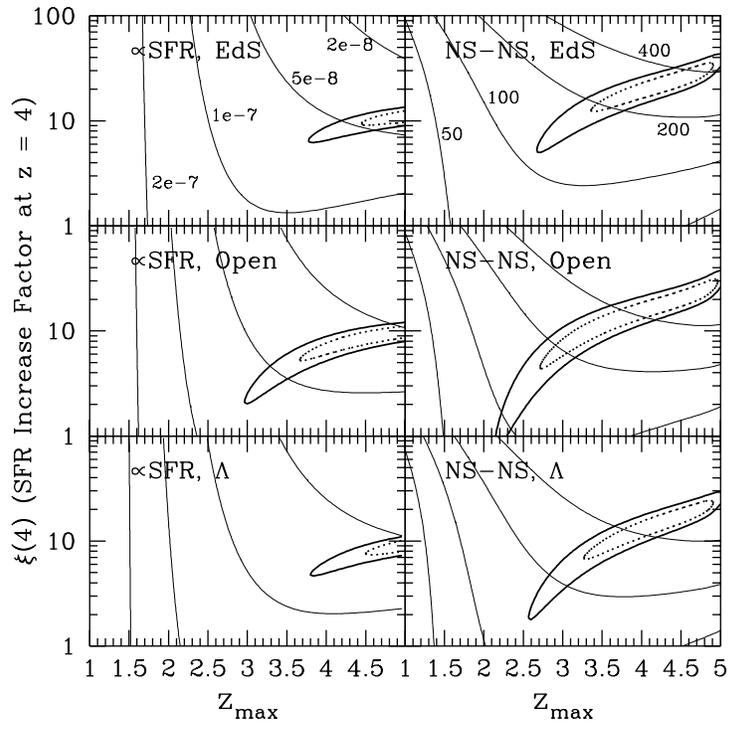,width=10cm}
  \end{center}
\caption{The same as Fig. \protect\ref{fig:zmax-SFR}, but $\xi(1)=3$.}
\label{fig:zmax-SFR-f3}
\end{figure}

However, as discussed in \S 3, some independent
estimates of $\xi(1)$ give lower values as low as $\xi(1) \sim 4$, and
there are acceptable models with such values of $\xi(1)$.
In order to see the allowed region about $\zmax$ in this case, let us consider
a statistical test under the condition that a particular value of $\xi(1)$
is given.
For a fixed value of $\xi(1) = 4$, we have searched the best-fit values of
$(\zmax, \xi(4))$ and results are shown in
Table \ref{table:zmax-SFR} for the long GRBs with six model parameter sets
(proportional or NS-NS, and the three cosmological models).
The goodness-of-fit is acceptable except for the proportional model
in the EdS and $\Lambda$-dominated universes, 
and allowed regions of $(\zmax, \xi(4))$
for the acceptable models are shown in Fig. \ref{fig:zmax-SFR} by thick
contours. These contours are defined by the increment of $\chi^2$ 
relative to $\chi^2_{\min}$ in Table \ref{table:zmax-SFR} with two degrees of
freedom. We have also repeated the same analysis but with $\xi(1) = 3$,
with which all the six models are acceptable (see Table \ref{table:zmax-SFR}), 
and the results are given in Fig. \ref{fig:zmax-SFR-f3}. The NS-NS
model gives a better fit than the proportional model because of 
the flatter evolution due to the time delay.
The maximum redshift is at least larger than
2 and likely larger than 3, depending on $\xi(4)$. 
If the distance scale of GRBs is determined
independently, the cosmic SFR at high redshift ($z \sim 4)$ can be
estimated.

We have also calculated the best-fit
normalization parameters, i.e., production rate of GRBs per unit mass
of star formation (events/$M_\odot$) for the proportional model
and the beaming factor ($4\pi/\Delta \Omega)$ for the NS-NS model,
for these acceptable models.
The results are shown by the thin contours in Figs. \ref{fig:zmax-SFR}
and \ref{fig:zmax-SFR-f3}. 
The production rate of GRBs required for the proportional model
is $\sim 5 \times 10^{-8}$ events$/M_\odot$ if the emission is isotropic, 
and the NS-NS model
requires a beaming factor of $\sim$ 200--400.

\subsection{Long and Short GRBs}
\begin{figure}
  \begin{center}
    \leavevmode\psfig{figure=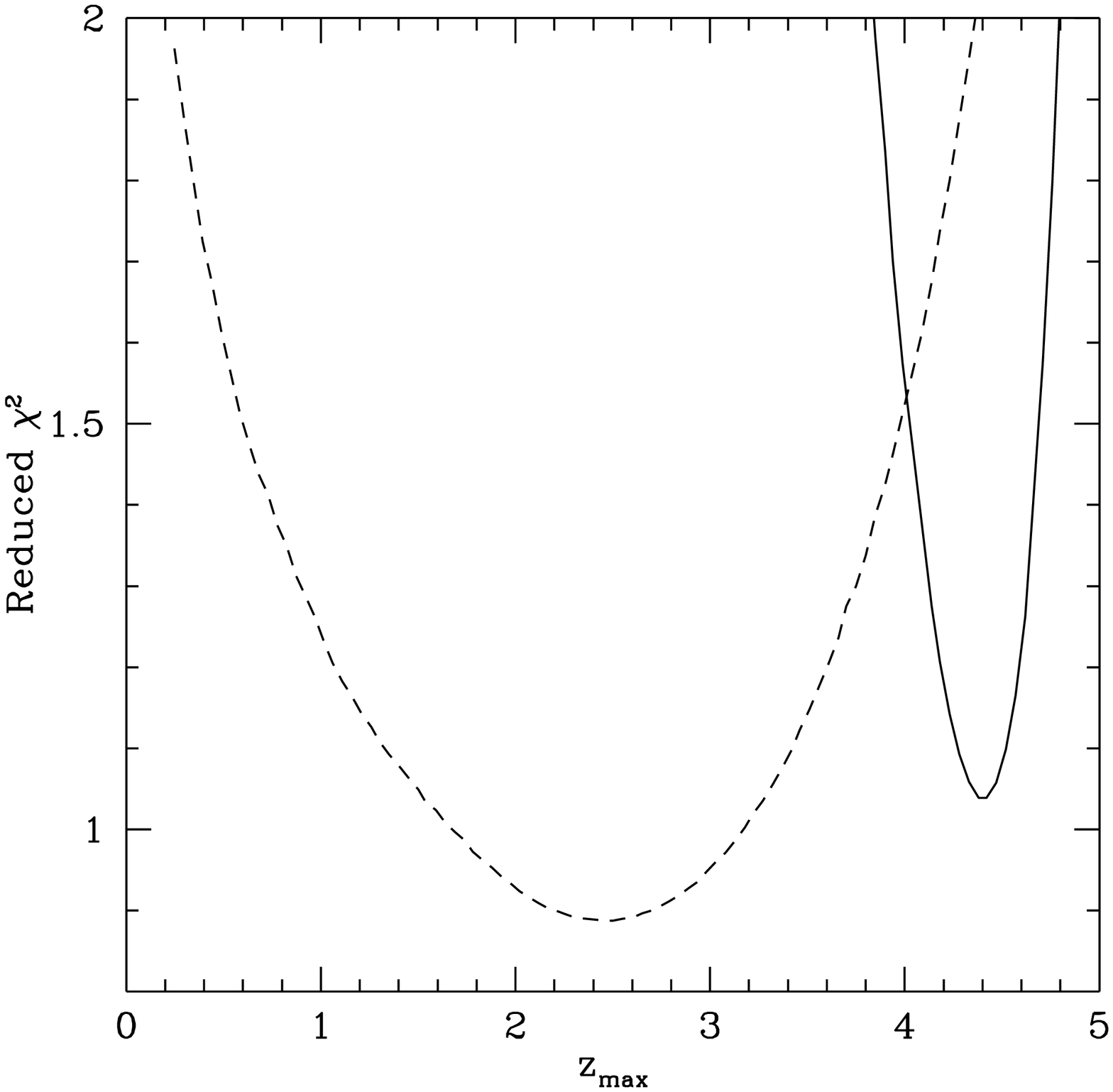,width=10cm}
  \end{center}
\caption{The reduced $\chi^2$ of the fit to the BATSE data as a function
of $\zmax$, with a fixed
cosmic SFR evolution $[(\xi(1), \xi(4)) = (4, 34.3)]$ in the NS-NS model.
The solid line is for the long GRBs and the dashed line for the short GRBs.
The Einstein-de Sitter universe with $h=0.5$ is assumed.}
\label{fig:chi-zmax}
\end{figure}

Because of the poor statistics of the short GRBs, we cannot derive strong
constraints on them.
However, it may be interesting to see whether there is a difference
between the distance scale of the long and short GRBs, provided that
the two populations are originated by the same physical event, i.e., 
the GRB rate evolution of the two populations as a function of $z$
is the same. In Figure 
\ref{fig:chi-zmax} we plotted the reduced $\chi^2$ of the fit
as a function of $\zmax$ for the long and short GRBs in the NS-NS model,
using a fixed star formation history [$(\xi(1), \xi(4)) = (4, 34.3)$].
This result suggests that $\zmax$ of the long GRBs is larger than
that of the short GRBs. Combined with the result of Fig. 
\ref{fig:luminosity}, this result confirms that 
the total energy emitted by the short GRBs 
in the restframe 50--300 keV is more than one order of magnitude
smaller than that by the long GRBs if the beaming factor of these
two populations is the same. 

It may be rather interesting to note that, although the total energy is
different for the two populations by a factor of more than 10, the peak
luminosity of the two populations is remarkably similar (see Fig. 
\ref{fig:luminosity}). There may exist a physical mechanism which
regulates the peak luminosity to a certain value in the shock dissipation of
relativistic kinetic energy, and it may give a hint to understand the origin
of the two populations.

\section{Discussion}
As shown in this paper, the star formation history in the universe
estimated by UV observations is completely inconsistent with
the brightness distribution of GRBs if the standard candle approximation
is valid. The origin of the discrepancy is too steep evolution
of SFR in the relatively low redshift range of $z$ = 0--1.
It is of course an option that GRBs are not related to death of massive stars,
but considering that most of theoretical GRB models are associated to 
collapse of massive stars, possibilities of reconciling the discrepancy
under the condition that GRB rate is related to SFR should be considered.

Clearly one possibility is intrinsic luminosity dispersion of GRBs. 
Krumholz, Thorsett, \& Harrison (1998) suggested that
luminosity dispersion by a factor of more than 20 can reconcile the
UV estimated SFR evolution and GRB distribution, assuming a power-law
distribution of GRB luminosity. However, although some luminosity dispersion
is likely as suggested from diversity of observational properties of GRBs,
it is not clear that the luminosity dispersion is actually such a broad one.
Krumholz et al. (1998) claimed the redshift estimations for GRB970508
($z=0.835$; Metzger et al. 1997) and 
GRB971214 ($z = 3,42$; Kulkarni et al. 1998) as an evidence
of broad luminosity dispersion. However, as is well known, the absorption
line at $z=0.835$ gives only an lower limit for the redshift of GRB970508.
The identification of the host galaxy for GRB971214 is also not perfect
but there is a doubt
that the `host' galaxy is accidentally on the line of sight. 
The shape of distribution of GRB luminosity function is also
unknown. The luminosity dispersion of GRBs is a possibility to reconcile the
discrepancy but it requires further observational confirmation.

On the other hand, it has recently been discussed that 
UV-estimated star formation history may be significantly biased due to
neglecting the effect of interstellar dust extinction 
(Meurer et al. 1997; Pettini et al. 1997; Sawicki \& Yee 1998;
Heckman 1998; Cimatti et al. 1998), as discussed
in \S 3. Therefore the possibility that
the strong SFR evolution in $z$ = 0--1 inferred from UV observations
is overestimated should be considered. Hammer et al. (1997, hereafter H97) 
pointed out that CFRS galaxies at $z \sim 1$ 
show signatures of low metallicity and smaller dust extinction,
which may cause an overestimate of SFR evolution in $z$ = 0--1.
H97 made a simple model to include this effect,
in which the conversion factor between
SFR and [OII] luminosity depends on the restframe $(U-V)$ color of
galaxies and the dependence on the color is determined by SFR calibration
of local galaxies. Then they found that the SFR evolution is $\xi(1)
\sim 6.9$ from the CFRS data in the Einstein-de Sitter universe, 
which is considerably lower than the pure evolution of
2800 {\AA} or [O II] comoving luminosity density
($\xi(1) \sim 15$).
If we change the cosmological model into the open [($\Omega_0, \Omega_\Lambda)
= (0.2, 0)$] or the $\Lambda$-dominated universe
[($\Omega_0, \Omega_\Lambda) = (0.2, 0.8)$], the value of 
UV-estimated $\xi(1)$ is further reduced to $\xi(1) = $
5.33 and 3.81, respectively. These values are now near the
independent estimate ($\xi(1) \sim 4)$
based on galaxy evolution models which
reproduce the present-day properties of galaxies (Totani, Yoshii, \& Sato
1997). Therefore we conclude that more modest evolution of SFR in
$z $ = 0--1 is not unreasonable and this would significantly remedy the
discrepancy between UV observations and GRB distribution.

\section{Summary and Conclusions}
In this paper we have presented a detailed study on the possible relation 
between the brightness distribution of gamma-ray bursts  
and the cosmic star formation history. The long and short GRBs in the 4B 
BATSE catalog are analyzed separately. We proposed a new method to determine
the average GRB spectrum in which the curvature of GRB spectra compared to
a pure power-law is taken into account based on
the observed $\alpha$-$P$ correlation.
Various models of the cosmic star formation history 
are tried considering
the present uncertainties in the observational
estimate of SFRs,
and implications of the GRB distribution on the star formation history
and galaxy evolution were discussed. 

We have shown that the evolution of 
SFR in $z$ = 0--1 is crucially important for the fit of GRB 
brightness distribution, and SFR evolution in this range is strongly
constrained if the standard candle approximation is valid.
The analysis on the long GRBs suggests that,
in the Einstein-de Sitter universe
with $(h, \Omega_0, \Omega_\Lambda) = (0.5, 1, 0)$, the SFR
increase factor from $z$ = 0 to 1 [$\xi(1)$] should be smaller than 3.0
(95 \% C.L.) if the GRB rate is proportional to SFR, and than
3.8 if the GRBs are produced by binary neutron-star mergers.
For short GRBs, $\xi(1)$ is constrained as $\xi(1) < 4.5$
and $\xi(1) < 10$ for the proportional and NS-NS models, respectively.
These values are significantly smaller than the current
estimate ($\xi(1) \sim$ 15; Lilly et al. 1996) of SFR evolution 
based on the UV luminosity density, but marginally 
consistent with a theoretical
estimate ($\xi(1) \sim 4$) by galaxy evolution models based on
the local properties of galaxies (Totani, Yoshii, \& Sato 1997). 
These results are consistent
with Totani (1997) in which the star formation history based on the galaxy 
evolution model gave a good fit while the observational history
did not. 

We have discussed some possibilities to reconcile this
apparent discrepancy between the UV-estimated SFR evolution and GRB 
distribution, under the condition that GRBs are related to death of
massive stars. One possibility is intrinsic luminosity dispersion
of GRBs, although neither the width of dispersion nor the shape of
distribution is well known.
We have also argued that the uncomfortably large $\xi(1)$
inferred from the UV observation
may be an overestimation and the real value could be as low as about 4
if we choose different sets of cosmological parameters and/or
take account of the evolution of metallicity and dust extinction.
Therefore the UV observation itself can also be consistent with the
BATSE data and the galaxy evolution model.

For the case of low values of $\xi(1) \sim 4$, we have obtained some
constraints on the distance scale of GRBs and production rate of GRBs.
We could not find any acceptable fit for the long GRBs with $\zmax < 2$.
Therefore $\zmax$ is at least larger than 2 and 
likely in the range of $\zmax$ = 3--5, if the long GRBs are associated
to death of massive stars or NS-NS mergers.
The production rate of GRBs from star formation is $\sim
5 \times 10^{-8}$ [$M_\odot^{-1}$]
if GRB rate is proportional to SFR, and the beaming factor required
for the NS-NS merger scenario is about a few hundreds.

\begin{figure}
  \begin{center}
    \leavevmode\psfig{figure=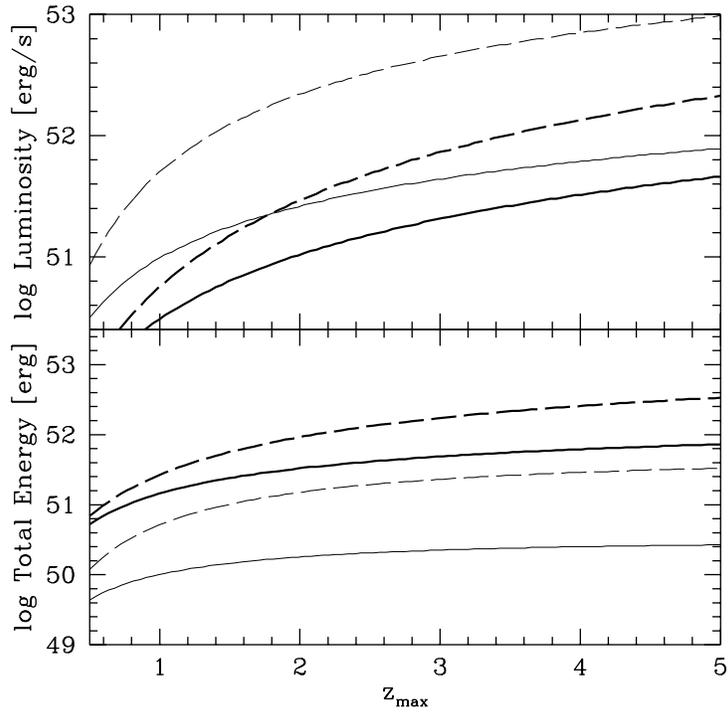,width=10cm}
  \end{center}
\caption{Peak luminosity and total energy emitted by GRBs, as functions of
the fitting parameter, $\zmax$. The thick lines are for the
long GRBs, in the energy range of 50--300 keV (solid) and 50--2000 keV
(dashed) at the rest frame, while the thin lines are for the short GRBs.
Isotropic emission is assumed in this figure.
}
\label{fig:luminosity}
\end{figure}

The maximum redshift of the short GRBs seems smaller than that of the long
GRBs if the GRB rate evolution is the same for the two populations. 
This confirms that the total energy emitted by the short bursts 
is smaller than that by the long bursts by more than one order of magnitudes
(see Fig. \ref{fig:luminosity}), if the beaming factor is the same.
On the other hand, the peak luminosity of the two populations
is remarkably similar ($\sim 2 \times 10^{51} \ \rm erg \ sec^{-1}$
in 50--300 keV). This might suggest that there is a physical mechanism
which regulates the energy loss rate of shock heated matter
in shocks generated by relativistic motion, in a wide range of 
total energy liberated as a relativistic fireball.

The author would like to thank an anonymous referee for careful reading
of this manuscript and useful comments.
He has been supported by the Research Fellowships of the Japan
Society for the Promotion of Science for Young Scientists, and 
the Grant-in-Aid for the
Scientific Research Fund (No. 3730) of the Ministry of Education, Science,
and Culture of Japan.

%%%%%%%%%%%%%%%%%%%% TABLES %%%%%%%%%%%%%%%%%%%%%%%%%%%%%%%%%%%
\begin{table}
%\scriptsize
%\footnotesize
\begin{center}
\caption{Best-fit Parameters for $\zmax$, $\xi(1)$, and $\xi(4)$}
\label{table:SFR-SFR}
\begin{tabular}{cccccccc}
\hline  \hline Cosmology\tablenotemark{\it a}
& Duration & Model\tablenotemark{\it b}
& $\zmax$ & $\xi(1)$ & $\xi(4)$ & $\chi^2_{\min}$ &
S.L. (\%)\tablenotemark{\it c} \\
\hline
EdS & long & prop. & 3.7 & 1.7 & 4.6 & 2.34 & 80 \\
EdS & long & NS-NS & 4.8 & 1.6 & 13.8 & 0.85 & 97 \\
EdS & short & prop. & 4.2 & 1.8 & 10.0 & 5.48 & 36 \\
EdS & short & NS-NS & 4.8 & 1.0 & 12.6 & 4.87 & 43 \\
Open & long & prop. & 2.9 & 1.7 & 2.1 & 2.00 & 85 \\
Open & long & NS-NS & 4.7 & 2.5 & 18.2 & 1.22 & 94 \\
$\Lambda$ & long & prop. & 2.5 & 1.0 & 1.0 & 2.05 & 84 \\
$\Lambda$ & long & NS-NS & 4.8 & 1.4 & 7.2 & 1.16 & 95 \\
\hline \hline
\end{tabular}
\end{center}
\tablenotetext{a}{Cosmological parameters used are $(h, \Omega_0, 
\Omega_\Lambda)$ = (0.5, 1, 0), (0.6, 0.2, 0), and (0.7, 0.2, 0.8)
for EdS, Open, and $\Lambda$, respectively.}
\tablenotetext{b}{The proportional model in which GRB rate is proportional
to SFR and the NS-NS model in which GRBs are produced by neutron star mergers.}
\tablenotetext{c}{The degree of freedom is $9-4 = 5$.}
\end{table}

\begin{table}
%\scriptsize
%\footnotesize
\begin{center}
\caption{Best-fit Parameters for $\zmax$ and $\xi(4)$. 
(Only for the long GRBs and $\xi(1)$ is fixed.)} 
\label{table:zmax-SFR}
\begin{tabular}{ccccccc}
\hline  \hline $\xi(1)$ 
& Model\tablenotemark{\it a} & Cosmology\tablenotemark{\it b}
& $\zmax$ & $\xi(4)$ & $\chi^2_{\min}$ &
S.L. (\%)\tablenotemark{\it c} \\
\hline
4 & prop. & EdS & 5.0 & 15.9 & 19.4 & 0.35 \\
4 & prop. & Open & 5.0 & 14.5 & 6.0 & 42 \\
4 & prop. & $\Lambda$ & 5.0 & 13.1 & 20.4 & 0.23 \\
4 & NS-NS & EdS & 4.5 & 36.3 & 8.3 & 22 \\
4 & NS-NS & Open & 4.6 & 31.6 & 2.3 & 89 \\
4 & NS-NS & $\Lambda$ & 4.4 & 24.0 & 9.91 & 13 \\
3 & prop & EdS & 5.0 & 11.0 & 7.9 & 25 \\
3 & prop & Open & 4.9 & 9.6 & 3.0 & 80 \\
3 & prop & $\Lambda$ & 5.0 & 9.1 & 9.97 & 13 \\
3 & NS-NS & EdS & 4.5 & 25.1 & 4.1 & 67 \\
3 & NS-NS & Open & 4.6 & 21.9 & 1.4 & 97 \\
3 & NS-NS & $\Lambda$ & 4.5 & 17.4 & 5.6 & 47 \\
\hline \hline
\end{tabular}
\end{center}
\tablenotetext{a}{The proportional model in which GRB rate is proportional
to SFR and the NS-NS model in which GRBs are produced by neutron star mergers.}
\tablenotetext{b}{Cosmological parameters used are $(h, \Omega_0, 
\Omega_\Lambda)$ = (0.5, 1, 0), (0.6, 0.2, 0), and (0.7, 0.2, 0.8)
for EdS, Open, and $\Lambda$, respectively.}
\tablenotetext{c}{The degree of freedom is $9-3 = 6$.}
\end{table}

%%%%%%%%%%%%%%%%%%%%%%%%%%%%%%%%%%% references

%%%%%%%%%%%%%%%%%%%%%%%%%%%%%%%%%%% figures
%\newpage

\begin{thebibliography}{200}
\bibitem[]{}
Arimoto, N. \& Yoshii, Y. 1987, \aap , 173, 23

\bibitem[]{}
Arimoto, N., Yoshii, Y., \& Takahara, F. 1992, A\&A, 253, 21 (AYT)

\bibitem[]{} 
Blinnikov, S. I. et al. 1984, Sov. Astr. Lett. 10, 177

\bibitem[]{}
Che, H., Yang, Y., Wu, M., \& Li, Q.B. 1997, ApJ, 483, L25

\bibitem[]{}
Cimatti, A., Andreani, P., R\"{o}ttgering, H., \& Tilanus, R. 1998,
Nature 392, 895

\bibitem[]{}
Connolly, A. J. et al. 1997, ApJ, 486, L11

\bibitem[]{}
Dermer, C. D. 1992, \prl, 68, 1799

\bibitem[]{} 
Faber, S. M. \&
Gallagher, J. S. 1979, \araa , 17, 135

\bibitem[]{}
Fenimore, E.E. and Bloom, J.S. 1995, ApJ, 453, 25

\bibitem[]{}
Gallego, J., Zamorano, J., Aragon-Salamanca, A., \& Rego, M. 1995,
ApJ, 455, L1 (erratum: 459, L43)

\bibitem[]{} 
Hammer, F. et al. 1997, ApJ, 481, 49 (H97)

\bibitem[]{}
Heckman, T.M. 1998, To appear in the proceedings of 
`The Most Distant Radio Galaxies', ed. D. Reidel
(astro-ph/9801155)

\bibitem[]{}
Horack, J. M., Mallozzi, R. S., \& Koshut, T. M. 1996, \apj, 466, 21

\bibitem[]{}
Kouveliotou, C. et al. 1993, ApJ, 413, L101

\bibitem[]{}
Kouveliotou, C. et al. 1997, IAU Circ. 6660

\bibitem[]{} 
Kulkarni, S. R. et al. 1998, Nature, 393, 35

\bibitem[]{}
Krumholz, M., Thorsett, S.E., \& Harrison, F.A. 1998, preprint,
astro-ph/9807117

\bibitem[]{}
Lilly, S. J., F\`{e}vre, O. Le., Hammer, F., \& Crampton, D. 1996,
ApJ, 460, L1

\bibitem[]{}
Lipunov, V. M. et al. 1995, \apj, 454, 593

\bibitem[]{}
Madau, P., Pozzetti, L., \& Dickinson, M. 1998, ApJ, 498, 106

\bibitem[]{}
Mallozzi, R. S., Pendleton, G. N., \& Paciesas, W. S. 1996,
\apj, 471, 636 

\bibitem[]{}
Mao, S. \& Paczy\'{n}ski, B. 1992, \apj, 388, L45

\bibitem[]{}
Meegan, C. A. et al. 1992, Nature, 355, 143

\bibitem[]{}
Meegan, C. A. et al. 1996, \apjs, 106, 65

\bibitem[]{}
Metzger, M.R. et al. 1997, Nature, 387, 878

\bibitem[]{}
Meurer, G.R. et al. 1997, AJ, 114, 54

\bibitem[]{}
Norris, J.P. et al. 1994, ApJ, 424, 540

\bibitem[]{}
Norris, J.P. et al. 1995, ApJ, 439, 542

\bibitem{}
Paciesas, W. S. et al. 1997, 4B BATSE catalog is available at
http://cossc.gsfc.nasa.gov/cossc/BATSE.html

\bibitem[]{}
Petrosian, V. \& Lloyd, N.M. 1997, To appear in the proceedings of 
the 4th Huntsville Gamma-Ray Burst Symposium, eds. C.A.Meegan, P.Cushman
(astro-ph/9711193)

\bibitem[]{}
Pettini, M. et al. 1997, to appear in `The Ultraviolet Universe at Low and High
Redshift', ed. W. Waller, (Woodbury: AIP Press), astro-ph/9708117

\bibitem[]{}
Piran, T. 1992, \apj, 389, L45

\bibitem[]{}
Portegies Zwart, S.F. \& Yungelson L.R. 1998, A\&A, 332, 173

\bibitem[]{}
Press, W.H. et al. 1992, Numerical Recipes, 2nd edition 
(Cambridge: Cambridge Univ. Press)

\bibitem[]{}
Sahu, K. et al. 1997, ApJ, 489, L127

\bibitem[]{}
Sawicki, M \& Yee, H.K.C. 1998, AJ, in press (astro-ph/9712216)

\bibitem[]{}
Totani, T. 1997, ApJ, 486, L71 (T97)

\bibitem[]{}
Totani, T., Yoshii, Y., \& Sato, K. 1997, ApJ, 483, L75 (TYS)

\bibitem[]{} 
Wijers, R.M.J., Bloom, J.S., Bagla, J.S. \& Natarajan, P. 1998, MNRAS,
294, L17

\end{thebibliography}
\end{document}